# Dynamically-generated pure spin current in single-layer graphene

Zhenyao Tang[1,#], Eiji Shikoh[1,#], Hiroki Ago[2], Kenji Kawahara[2], Yuichiro Ando[1], Teruya Shinjo[1] and Masashi Shiraishi[1,*]

1. Graduate School of Engineering Science, Osaka Univ. Toyonaka 560-8531, Japan
2. Institute for Material Chemistry and Engineering, Kyushu Univ. Fukuoka 816-8508, Japan

# These two authors contributed equally to this work.
*Corresponding author: shiraishi@ee.es.osaka-u.ac.jp

**Abstract**

The conductance mismatch problem limits the spin-injection efficiency significantly, and spin-injection into graphene has been usually requiring high-quality tunnel barriers to circumvent the conductance mismatch. We introduce a novel approach, which enables generation of a pure spin current into single-layer graphene (SLG) free from electrical conductance mismatch by using dynamical spin injection. Experimental demonstration of spin-pumping-induced spin current generation and spin transport in SLG at room temperature was successfully achieved and the spin coherence was estimated to be 1.36 μm by using a conventional theoretical model based on Landau-Lifshitz-Gilbert equation. The spin coherence is proportional to the quality of SLG, which indicates that spin relaxation in SLG is governed by the Elliot-Yafet mechanism as was reported.

Electrical spin injection and generation of a pure spin current in graphene using non-local electrical technique has opened a new frontier in molecular spintronics [1-3]. A number of studies have been performed using this method, investigating spin drift [4,5], long spin coherence in bilayer graphene (BLG) [6,7] and robustness of spin polarization [8]. Since SLG possesses a linear band structure around the K and K' points, the effective mass of spin carriers is quite small, which allows us to expect much better spin coherence than that observed in BLG. However, the experimentally observed coherence in SLG is still limited and shorter than that in theoretical predictions. In addition, the spin relaxation mechanisms in SLG and BLG are reported to be different, namely, Elliot-Yafet in SLG and D'yakonov-Perel in BLG [6]. Hence, there are still many issues in spin transport in graphene that need to be clarified, and the establishment of a novel technique for spin injection and generation of a pure spin current in graphene is strongly desired for discussing spin transport phenomena in graphene from a different and new standpoints. Here, we report on a novel approach to spin injection and generation of pure spin current in SLG enabling to circumvent the conductance mismatch problem [9], where dynamical spin injection without using electric current for generating a spin current is established. The estimated spin coherence at room temperature (RT) is 1.36 μm. This achievement provides a new platform for discussing spin transport physics in graphene.

The SLG used in this study was a large-area SLG grown by chemical vapor deposition (CVD). The SLG was transferred to an $SiO_2$/Si substrate ($SiO_2$ is 300 nm thick, and the details were described in elsewhere [10]). Figure 1a shows a schematic image of the sample used in this study. A $Ni_{80}Fe_{20}$ film (Py, 25 nm in thick and 900×300 μm² in size) and a Pd wire (5 nm in thick and 1 μm in width) were fabricated using electron beam lithography and an evaporation method. An Al capping layer was evaporated onto the Py layer in order to prevent oxidation of the Py. The gap length between the Py and the Pd was ca. 970 nm. An external magnetic field was applied by changing the angle $\theta$, as shown in Fig. 1(a) ($\theta$ is equal to 0 degree when the magnetic field is parallel to the SLG plane). Fig. 1(b) shows typical Raman spectra of the SLGs, and the Raman spectra were obtained by using a Raman spectrometer (Tokyo Instrument, Nanofinder30). We prepared two different qualities of the SLGs (sample #1 and #2). The quality of sample #1 is better than that of sample #2, since D-band can be observed only from sample #2 (see Fig. 1(b)). Furthermore, we clarified that they are SLGs by measuring transport characteristics (see supplemental information).

The principle of dynamical spin injection is as follows. Magnetization dynamics in a ferromagnet (the Py in this study) are described by the Landau-Lifsitz-Gilbert (LLG) equation as,

$$\frac{dM}{dt} = \gamma H_{eff} \times M + \alpha M \times \frac{dM}{dt}, \qquad (1)$$

where $\gamma$, $M$, $H_{eff}$, $\alpha$ and $M_s$ are the gyro-magnetic ratio of the ferromagnet, the time-dependent magnetization of the ferromagnet, an external magnetic field, the Gilbert damping constant and saturation magnetization of the ferromagnet, respectively. The first and the second terms are the field term that describes magnetization precession and the damping term that describes damping torque. The Ferromagnetic resonance (FMR) occurs when a microwave (9.62 GHz in this study) is applied to the ferromagnet. The damping torque is suppressed by the applied microwave under the FMR, which induces pumping of spins into SLG due to spin angular momentum conservation. The pumped spins induce spin accumulation in the SLG, which allows generation of a pure spin current in the SLG (see supplemental information). The propagating spins are absorbed in the Pd wire, where a pure spin current ($J_s$) is converted to a charge current ($J_c$) due to the inverse spin Hall effect (ISHE) [11] and an electromotive force at the Pd wire is generated. The ISHE is the reciprocal effect of the spin Hall effect, and the $J_c$ is described as $J_c \sim J_s \times \sigma$ ($\sigma$: spin direction). Note that the sign of $J_c$ is changed by varying the direction of spin ($\sigma$), namely, the sign of the electromotive force at the Pd wire becomes opposite, when the pure spin current is successfully generated and propagated in the SLG and $\sigma$ is reversed by a static external magnetic field. Figure 1(c) shows the FMR spectra of the Py with and without the SLG/Pd. The line width of the spectrum from the sample with the SLG/Pd is larger than that without the SLG/Py, which is attributed to the modulation of $\alpha$ due to successful spin pumping into the SLG/Pd.

Figures 2(a)-2(c) show the FMR signals as a function of $\theta$, where the FMR of the Py occurs in every condition. The electromotive force of the Pd wire is shown in Figs. 2(d)-2(f), where the electromotive force is observed when $\theta$ is set to be 0 and 180 degrees, whereas no signal was observed at $\theta$ = 90 degree. Since this finding is in accordance with the angular dependence of the electromotive force due to the ISHE ($J_c \sim J_s \times \sigma$), the observed electromotive force is ascribed to the ISHE of the Pd, which is due to spin pumping into the SLG and achievement of spin transport of a dynamically generated pure spin current at room temperature. Here, note that there is no spurious effect with the observed symmetry and only the ISHE possesses the symmetry. For example, the anomalous Hall effect (AHE) signal can be included as a spurious signal, which may impede detection of the ISHE signals, but the AHE does not show such an external magnetic field dependence of the electromotive force. The following investigations also support the result that the ISHE signals were dominantly observed. The theoretical fitting was performed in order to separate the ISHE and the AHE signals in the observed electromotive force by using the following equation, $V = V_{ISHE} \frac{\Gamma^2}{(H-H_{FMR})^2 + \Gamma^2} + V_{AHE} \frac{-2\Gamma(H-H_{FMR})}{(H-H_{FMR})^2 + \Gamma^2} + aH + b$, where $V_{ISHE}$ is the electromotive force, $V_{AHE}$ is the voltage due to the AHE, $H$ is an external static magnetic field for the FMR, $H_{FMR}$ is the magnetic field where the FMR occurs, and $\Gamma$, $a$, $b$ are fitting parameters. An example of the fitting is shown in Fig. 3(a), and $V_{ISHE}$ and $V_{AHE}$ are estimated to be $1.16 \times 10^{-5}$ V and $6.67 \times 10^{-7}$ V, respectively. The contribution from the AHE to the electromotive force by the ISHE was revealed to be very weak. Figure 3(b) shows the microwave power dependence of the electromotive force at the Pd for the Py/SLG/Pd sample. The electromotive force at the Pd wire, $V_{ISHE}$, is nearly proportional to the microwave power, which indicates that the density of the generated spin current in the SLG proportionally increases with the applied microwave power [12]. Also, as shown in Fig. 3(c), the electromotive force proportionally increased with the microwave power, although the voltage due to the AHE was small enough. In fact, the ratio of the signal intensities by the ISHE and AHE at 200 mW was estimated to be 17, which indicates that the ISHE signal is dominant in the observed electromotive force and that the observed signal was mainly due to spin transport in the SLG.

Spin coherence in the experiment is estimated based on a conventional spin pumping theory [13]. The dynamical magnetization process induces spin pumping from the Py layer to the graphene layer and generates a spin current, $j_s$, as,

$$j_s = \frac{\omega}{2\pi} \int_0^{2\pi/\omega} \frac{\hbar}{4\pi} g_r^{\uparrow\downarrow} \frac{1}{M_s^2} [M(t) \times \frac{dM(t)}{dt}]_z dt. \qquad (2)$$

Here, $g_r^{\uparrow\downarrow}$ and $\hbar$ are the real part of the mixing conductance [14] and the Dirac constant, respectively. Note that $g_r^{\uparrow\downarrow}$ is given by,

$$g_r^{\uparrow\downarrow} = \frac{2\sqrt{3}\pi M_s \gamma d_F}{g\mu_B \omega}(W_{Py/SLG} - W_{Py}), \qquad (3)$$

where $g$, $\mu_B$ and $d_F$ are the g-factor, the Bohr magneton and thickness of the Py layer, respectively, and, $d_F$, $W_{Py/SLG}$ and $W_{Py}$ in this study were 25 nm, 3.10 mT and 2.57 mT, respectively. From Eqs. (1)-(3), the spin current density at the Py/SLG interface is obtained as,

$$j_s = \frac{g_r^{\uparrow\downarrow}\gamma^2 h^2 \hbar[4\pi M_s \gamma + \sqrt{(4\pi M_s)^2 \gamma^2 + 4\omega^2}]}{8\pi\alpha^2[(4\pi M_s)^2 \gamma^2 + 4\omega^2]},$$ where $h$ is the microwave magnetic field,

set to 0.16 mT at a microwave power of 200 mW. As discussed above, the broadening of $W$ in the Py/SLG compared with that in the Py film was attributed to spin pumping into the SLG and $g_r^{\uparrow\downarrow}$ in the Py/SLG layer was calculated to be $1.6\times 10^{19}$ m$^{-2}$, and thus $j_s$ was calculated to be $7.7\times 10^{-9}$ Jm$^{-2}$. Here, half of the generated $j_s$ can contribute to the electromotive force in the Pd electrode in our device geometry, since a pure spin current diffuses isotropically. The generated $j_s$ decays by spin diffusion in SLG, $j_s$ decays to $j_s \cdot \exp(-\frac{d}{\lambda})$ when spins diffuses to Pd wire. Furthermore, the electromotive force taking the spin relaxation in the Pd wire into account can be written as, $V_{ISHE} = \frac{w\theta_{SHE}\lambda_{Pd}\tanh(d_{Pd}/2\lambda_{Pd})}{d_{SLG}\sigma_{SLG} + d_{Pd}\sigma_{Pd}}(\frac{2e}{\hbar})j_s$, in the simplest model. Here, $w$, $\lambda_{Pd}$, $d_{Pd}$ and $\sigma_{Pd}$ are the length of the Pd wire facing the Py (900 nm), the spin diffusion length (9 nm) [15], the thickness (5 nm) and the conductivity of the Pd, respectively, and $d_{SLG}$ and $\sigma_{SLG}$ are the thickness (assumed to be ca. 0.3 nm) and conductivity of the SLG (measured to be ca. $3.10\times 10^6$ S/m under the zero gate voltage application. See supplemental information). The spin-Hall angle in a Py/Pd junction, $\theta_{SHE}$, has been reported to be 0.01 [12], which allows us to theoretically estimate the electromotive force in the Pd wire as $2.37 \times 10^{-5}$ V if no spin relaxation occurred in the SLG. In contrast, the experimentally observed electromotive force was $1.16 \times 10^{-5}$ V, this discrepancy is ascribed to dissipation of spin coherence during spin transport in the SLG (the decay of $j_s$), which can be described as an exponential damping dependence on the spin transport. From the above calculations, the spin coherent length in the SLG is estimated to be 1.36 μm. For comparison, we carried out the same experiments by using the other sample (sample #2, the gap length was measured to be 780 nm.), of which quality is not as good as that of sample #1 (the conductivity was measured to be $6.40 \times 10^5$ S/m under the zero gate voltage application). Figures 4(a)-4(f) show the result, and here again, the obvious ISHE signals and the inversion of the ISHE signals can be seen as the external magnetic field was reversed. However, the signals were comparatively weak and the estimated spin coherent length was ca. 460 nm, which is in agreement with the sample qualities and also with the reported spin relaxation mechanism in SLG, i.e., the Elliot-Yafet type. These observations also corroborates that our result is attributed to dynamical spin injection and spin transport in the SLGs. The spin coherent length in the CVD-grown SLG, which was estimated by using the dynamical method, is good and comparable to the previously reported value (1.1 μm) estimated by using an electrical method [16]. In contrast, the spin coherence of the SLG in this study is much better than that in p-Si (ca. 130 nm) in our previous study [17], which directly indicates the strong superiority of graphene for spin transport.

In summary, we successfully demonstrated dynamical spin injection, resulting in generation of spin current in SLG at room temperature, which enables generation of a pure spin current free from electrical conductance mismatch. The spin coherent length of CVD-grown

SLG was 1.36 μm. This achievement provides a novel platform for discussing spin transport physics in SLG from a new viewpoint.

A part of this study was supported by the Japan Science and Technology Co. (JST), Global COE program of Osaka Univ., and the Japan Society for the Promotion of Science (JSPS).

**Figures and figure captions**

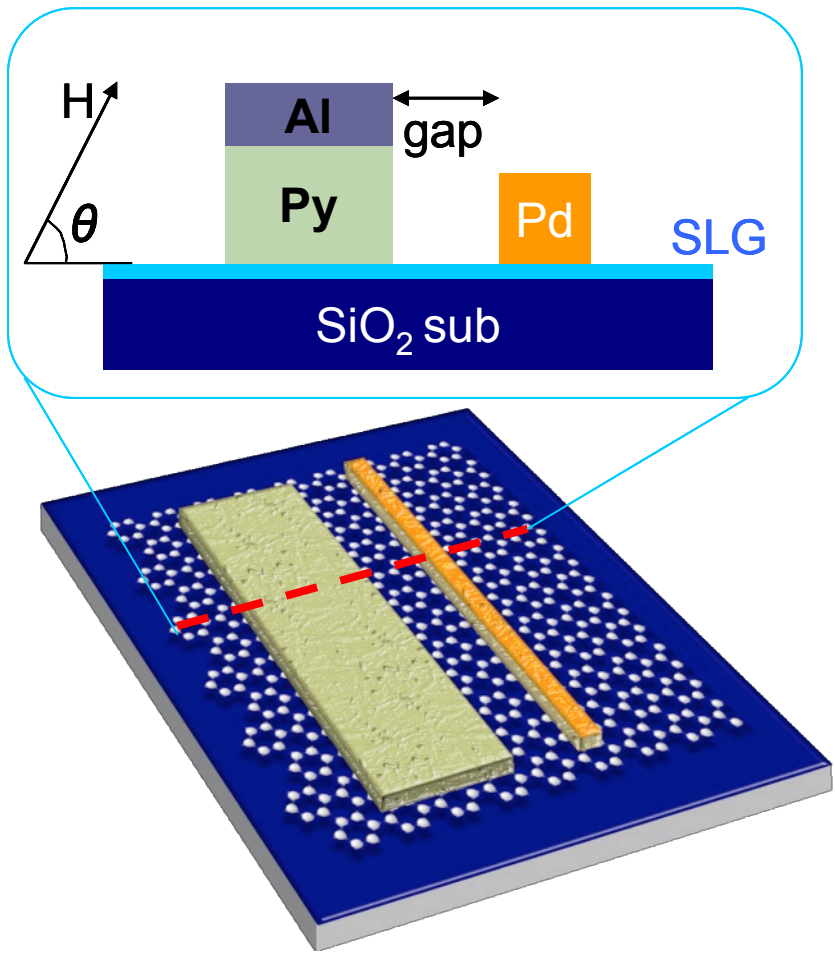

**Figure 1(a).** A schematic image of the SLG spin pumping sample. The CVD-grown SLG is transferred to the SiO$_2$/Si substrate, and the Al/Py and the Pd electrodes are separately evaporated on the SLG. The angle of the external magnetic field is shown in the inset.

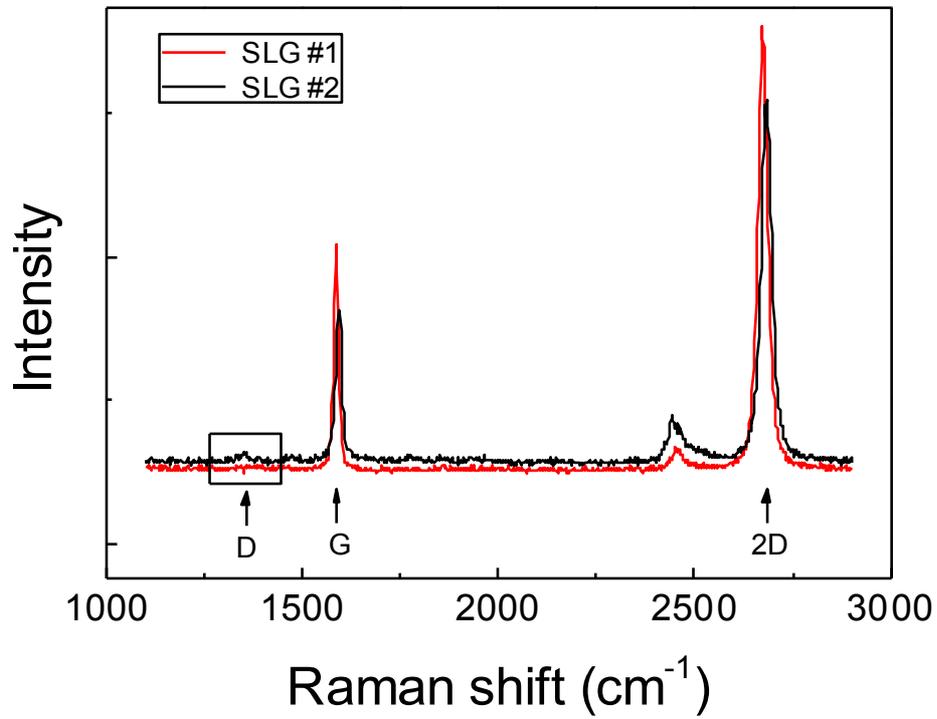

**Figure 1(b).** Raman spectra of the SLG just after the synthesis. The black and red solid lines are data from sample #1 and #2, respectively. The typical symmetric Raman peaks of the G- and 2D-bands from SLG can be seen. The D-band can be hardly seen in sample #1, which shows the defect of sample #1 is less than that of sample #2.

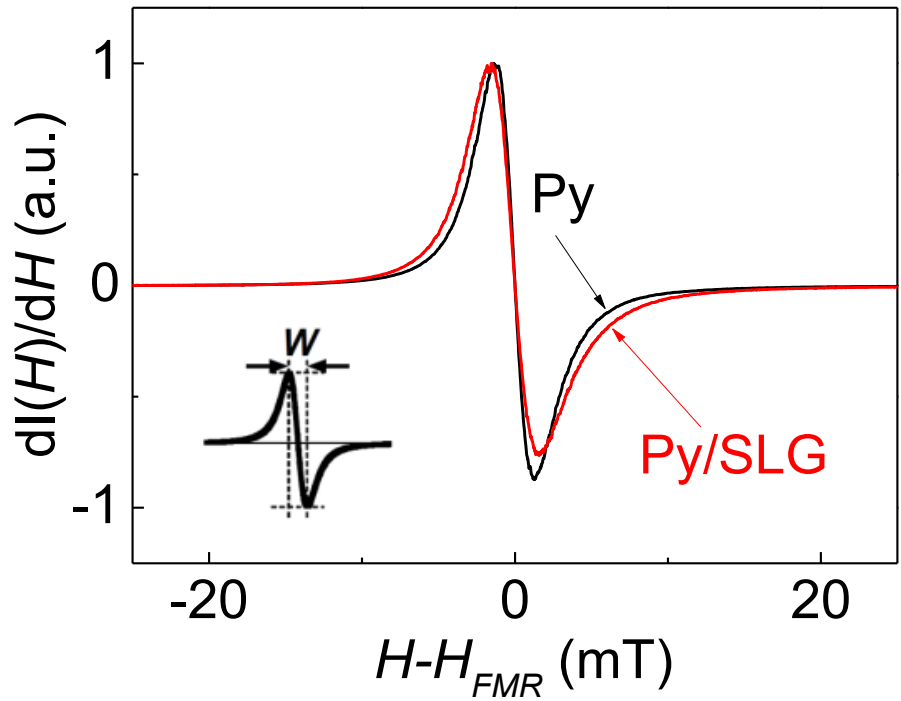

**Figure 1(c).** FMR spectra of the Py on the $SiO_2$ substrate (a black solid line) and the Py/SLG on the $SiO_2$ substrate (a red solid line, sample #1). An increase of the line width can be seen, which is attributed to a shift of the Gilbert damping constant, $\alpha$, namely, spin pumping into the SLG.

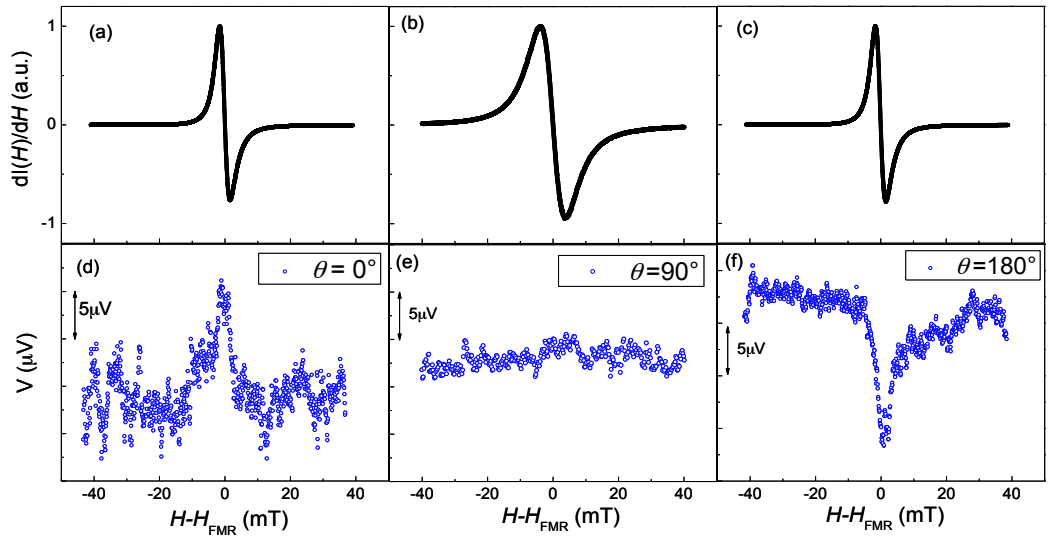

**Figure 2.** Results on spin pumping and spin transport in sample #1. (a)-(c) Ferromagnetic resonance spectra of the Py under the external magnetic field at (a) 0 degree, (b) 90 degree and (c) 180 degree under the microwave power of 200 mW. (d)-(f) Electromotive forces from the Pd wire on the SLG when $\theta$ is set to be (d) 0, (e) 90 and (f) 180 degrees under the microwave power of 200 mW. Electromotive forces are observed at $\theta$ = 0 and 180 degrees but the polarity of the signals are opposite, which is in accordance with the symmetry of the inverse spin Hall effect. No signal can be observed when the $\theta$ was set to be 90 degrees, which corroborates our observation originating from the spin pumping and propagation of a pure spin current in the SLG.

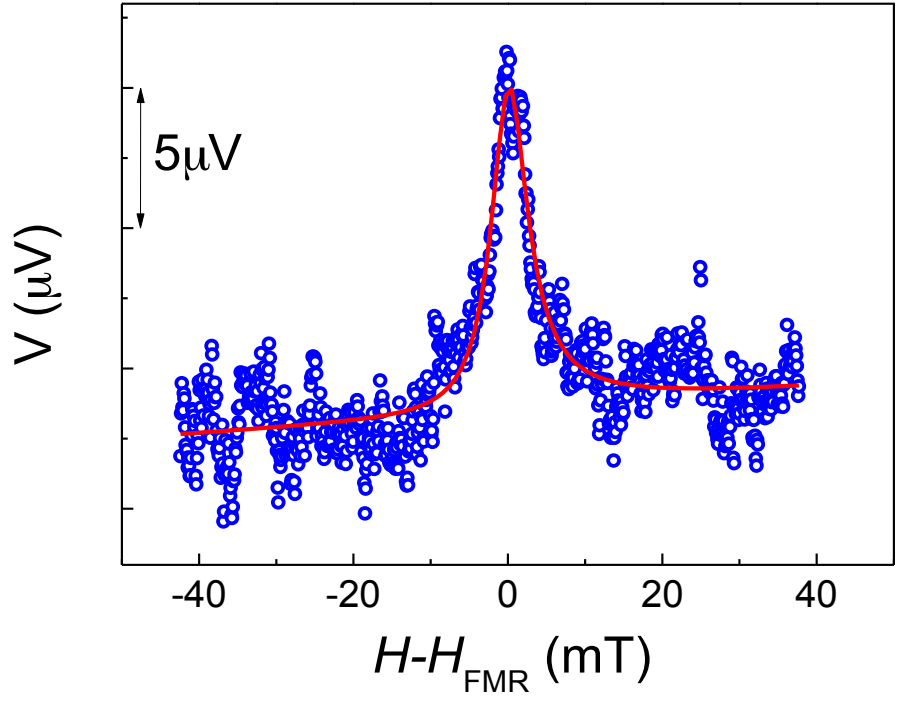

**Figure 3(a).** Results of the analysis of the contribution from the electromotive force due to ISHE and AHE. The open circles are external magnetic field dependence of $\Delta V$ for sample #1, where $\Delta V = [V(\theta = 0°) - V(\theta = 180°)]/2$ in order to eliminate the heat effect during measurement. The solid line is the fitting line, which is calculated by using Eq. (2).

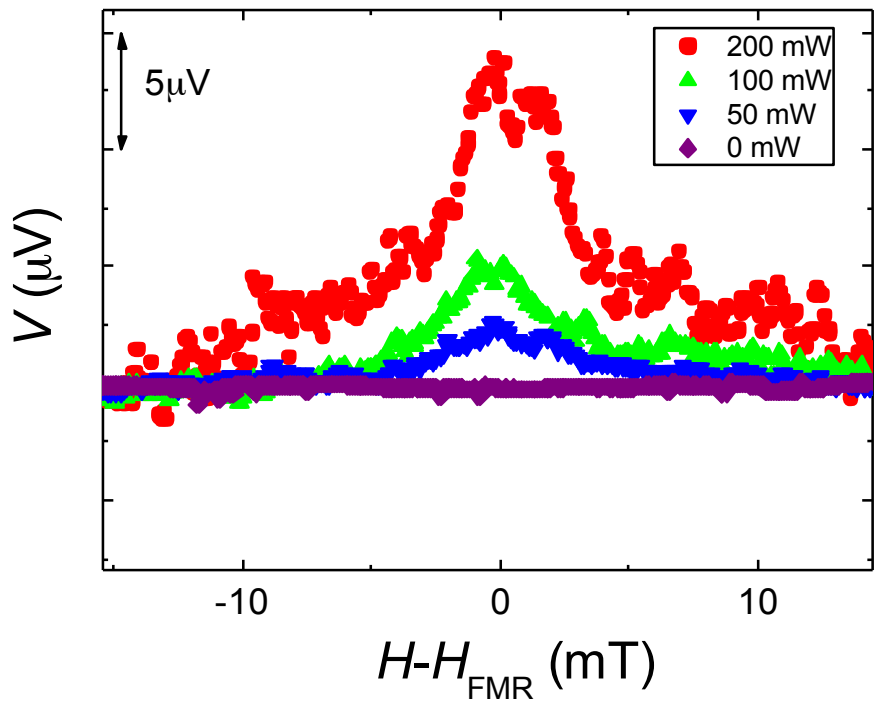

**Figure 3(b).** Microwave power dependence of the electromotive forces in the Pd wire. A magnetic field was applied parallel to the film plane. A monotonical increase of the electromotive force can be seen.

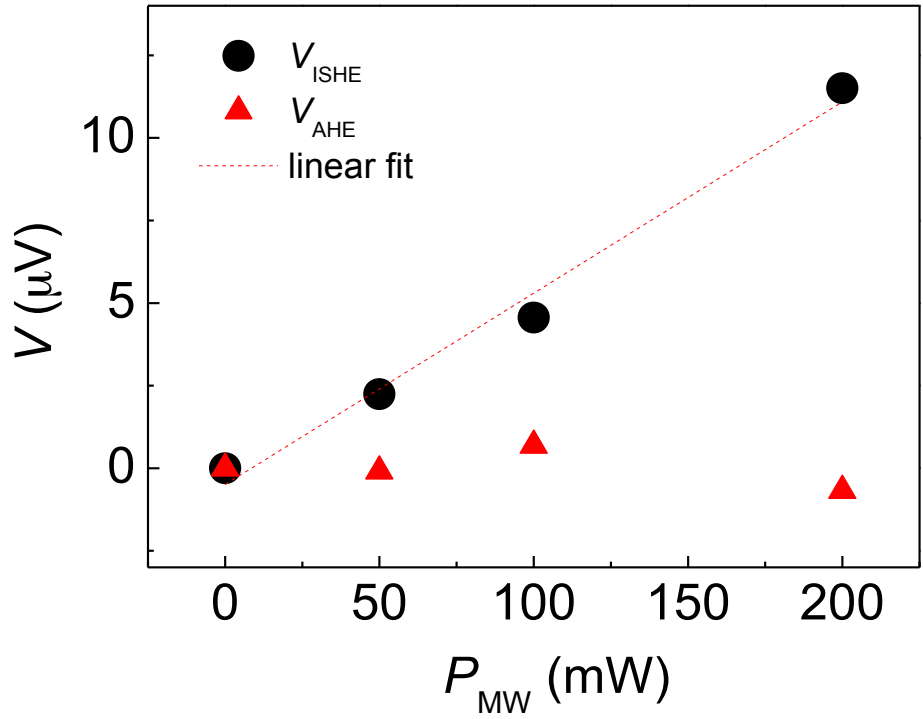

**Figure 3(c).** Microwave power ($P_{MW}$) dependence of $V_{ISHE}$ and $V_{AHE}$ measured for the Py/SLG/Pd sample. The contributions from the ISHE ($V_{ISHE}$) and anomalous Hall effect ($V_{AHE}$) are plotted by black circles and red triangles, respectively. The dashed line shows the linear fit of the data for the $V_{ISHE}$.

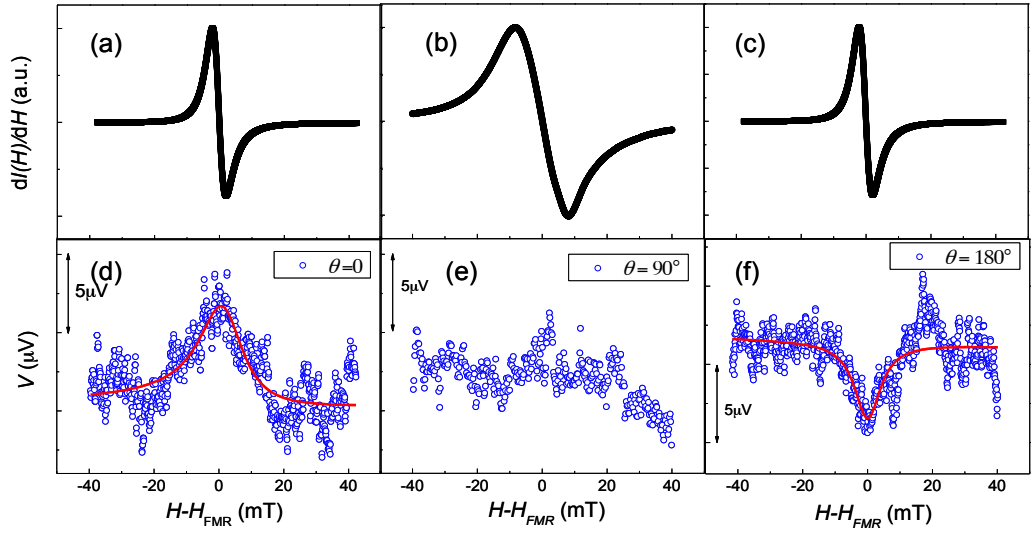

**Figure 4.** Results on spin pumping and spin transport in sample #2. (a)-(c) Ferromagnetic resonance of the Py under $\theta$ is set to be (a) 0 degree, (b) 90 degree and (c) 180 degree under the microwave power of 200 mW. (d)-(f) Electromotive forces from the Pd wire on the SLG when $\theta$ is set to be (d) 0, (e) 90 and (f) 180 degrees under the microwave power of 200 mW. Electromotive forces can be seen as expected, indicating successful dynamical spin injection. However, the signal was weak and noisy. The red solid lines in (d) and (f) show the fitting lines obtained by using Eq. (2).

**Supplemental Materials for "Dynamically-generated pure spin current in single-layer graphene" by Tang et al.**

In order to clarify that the graphene channel was definitely formed, we measured the field-effect transistor (FET) characteristics made of sample #1, and clarified that typical FET features were observed as shown in Fig. S1. Here, a slight distortion in the gate voltage dependence of source-drain current was observed, which often appears SLG-FETs with ferromagnetic contacts [S1]. This FET characteristic apparently indicates that the charge current flows in the SLG.

Figure S2 shows a schematic image of spin-pumping-induced generation of pure spin current and its isotropic diffusion in SLG. The isotropic spin diffusion is ascribed to the uniform electrochemical potential of spins in the SLG, since no electric field is applied to the SLG. The diffused pure spin current are adsorbed in the Pd wire and converted to an electric current due to ISHE.

[S1] R. Nouchi, M. Shiraishi, Y. Suzuki. *Appl. Phys. Lett.* **2008**, 93, 152104.

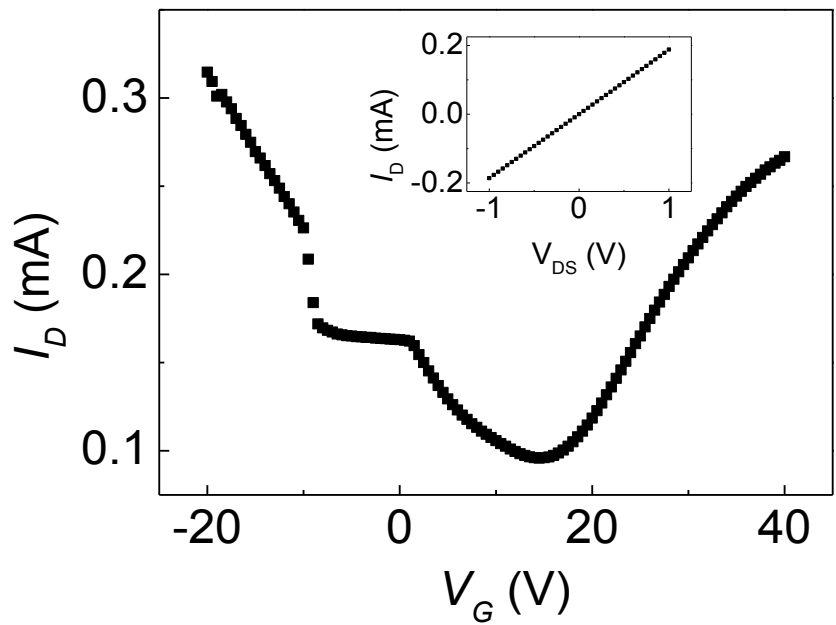

**Figure S1.** FET characteristics of SLG #1. The inset shows the the *I-V* curve of the source-drain current. The measurement was carried out at room temperature.

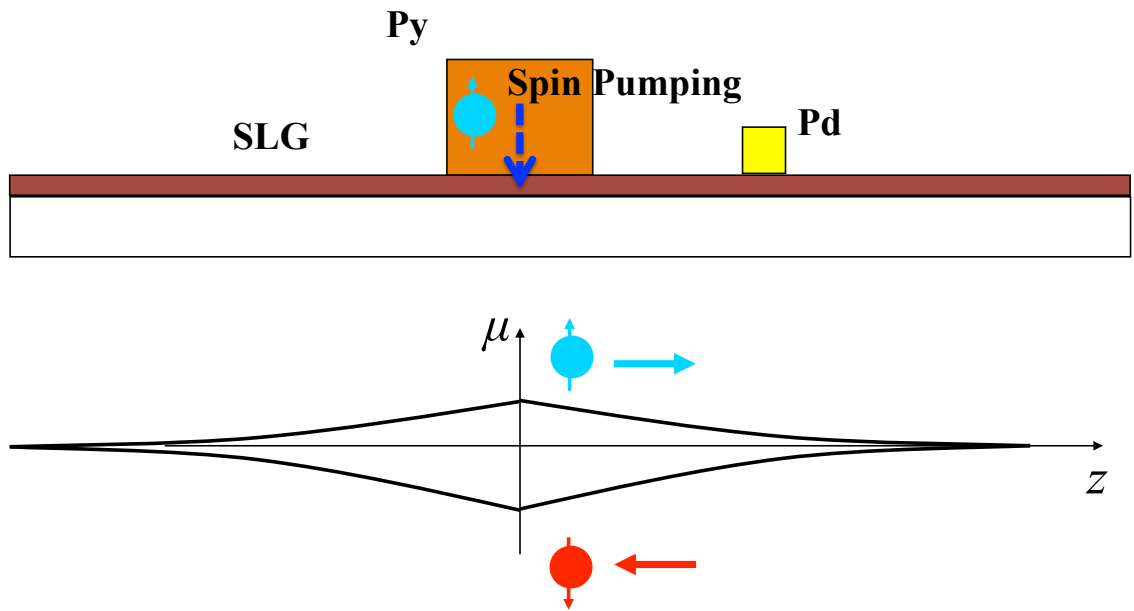

**Figure S1b.** A schematic image of spin pumping into the SLG. Spin angular momentum is transformed from the Py to the SLG due to π-d coupling, which induces spin accumulation in the SLG (the accumulation induces separation of electrochemical potentials, $\mu$, of the up- and down-spins). The accumulated spins diffuse in the SLG, resulting in a pure spin current. This pure spin current flows into the Pd electrode, where electromotive force arises due to the inverse spin Hall effect.